\documentclass[11pt]{mbd2013abstract}
\usepackage{graphicx}
\usepackage{amsmath} 

\usepackage[bf]{caption}

\newcommand{\p}{\mathcal{P}}

\usepackage{amsfonts}
\usepackage{amsmath}
\usepackage{color}

\newcommand{\ket}[1]{\left|#1\right\rangle}
\newcommand{\bra}[1]{\left\langle#1\right|}

\newcommand{\outm}[1]{\ket{#1}\bra{#1}}

\newcommand{\I}[1]{I^{(#1)}}

\begin{document}

%--- please fill-in information for your contribution --- 
%--- title, authors, afiliations ---

\title{Optimal Classical Hierarchical Correlation Quantification on Tripartite Mixed State Families}

\author{\presenter{Yuri Campbell}\ainst{1}, Jos\'{e} Roberto Castilho Piqueira\ainst{2}}

\address{%
  \ainstnum{1}Max-Planck-Institute for Mathematics in the Sciences, Inselstr. 22, 04103 Leipzig, Germany \url{campbell@mis.mpg.de}
  \ainstnum{2}Telecommunication and Control Engineering Department, Engineering School, University of Sao Paulo, Av. Prof. Luciano Gualberto, trv.3, n.158, Sao Paulo, SP ZIP:05508-900, Brazil
  \url{piqueira@lac.usp.br}%
  }

\Maketitle
\vspace{-.7cm}
\section*{Abstract} 
\vspace{-.3cm}
To which extent the whole of a system cannot be reduced into the sum of its parts? Apart from complexity theory, this question stands on the core of composite quantum states structure, given the intrinsic structural properties of Hilbert spaces' tensor product \cite{Bengtsson2006}.
This arises in both cases due to correlations playing a role in the overall observed dynamics, or statistical properties.

% There are at least a number of ways to formally define complexity. Most of them relate to some kind of minimal description of the studied object.
% Being this one in form of minimal resources of minimal effort needed to generate the object itself.
% This is usually achieved by detecting and taking advantage of regularities within the object.

% Regularities can commonly be described in an information-theoretic approach by quantifying the amount of correlation playing a role in the system, this being spatial, temporal or both.
% This is the approach closely related to the extent that the whole cannot be understood as only the sum of its parts, but also by their interactions.
% Feature considered to be most fundamental.
% Nevertheless, this irreducibility, even in the basic quantum informational setting of composite states, is also present due to the intrinsic structure of Hilbert spaces' tensor product \cite{Bengtsson2006}.

Those correlations give rise to regularities on those objects, being spatial, temporal or both.
As a way to study them, information-theoretic and minimal-description methods are commonly used.
Apart from the well known information measures, in the quantum setting there are also some proposed quantum informational minimal description-oriented approaches \cite{Mora2005}, based on a quantum extension of Kolmogorov's complexity.
Although in general among them a quantification is achieved, there is no clear structural characterization.
This being the main goal of the hierarchical estimation of correlations \cite{Kahle2009}, called here hierarchical information.
Those tools have been used in classical systems as Cellular Automata and Coupled Tent Maps \cite{Kahle2009}, and later formalized \cite{Ay11}.

%%%%%%%%%%%%%%%%%%%%%%%%%%%%%%%%%%%%%%%%%%%%%%%%%%%%%%%%%%%%%%%%%%%%%%%%%%%%%%%%%%%%%%%%%%%%%%%%%%%%%%%%%%%%%%%%%%%%%%%%%%%%%%%%%%%%%%%%%%%%%%

This is an approach essentially based on quantifying the degree of interaction from few to increasingly more parts of one system, in a fashion that as many parts contribute to the system dynamics, higher orders of interaction rise, giving room for richer dynamics, richer statistics.
Hierarchical information (HI) generalizes the concept of mutual information, giving it a structural interpretation.
% and multi-information, the later being a kind of mutual information among several random variables.
% Systems with higher interactions, have 
% This is done via the hierarchically construction of exponential families statistical models from the observed statistics.
% More formally, the hierarchical information is the generalization of both the concept of mutual information, the Kullback-Leibler (KL) divergence between the joint distribution and the product-distribution, and multi-information, which is a kind of mutual information among several random variables.
The hierarchical order of a model can be seen as the number of units interacting that actually contribute to the final statistics, those units are represented as random variables in the statistical model, and their interactions, as correlations.
%  in this case random variables in a joint distribution needed to fully describe the observed statistics via a exponential family model.
The hierarchical information then takes a set of random variables $S=\{X_i\},\; i=0,\cdots,N$, and computes for each statistical model of order $k$, $1\leq k\leq N$, the Kullback-Leibler divergence to the closest model of order $k-1$. Both models are constructed via exponential families.
In simple words, HI quantifies the degree to which the statistics of the $k-$th order model cannot be reduced to the $(k-1)-$th order one.
Building in this way a vectorial quantity $\I{k}$, measured in bits for each $k$, as it is a KL divergence.

%%%%%%%%%%%%%%%%%%%%%%%%%%%%%%%%%%%%%%%%%%%%%%%%%%%%%%%%%%%%%%%%%%%%%%%%%%%%%%%%%%%%%%%%%%%%%%%%%%%%%%%%%%%%%%%%%%%%%%%%%%%%%%%%%%%%%%%%%%%%%%

% Proposed by \cite{Kahle2009}, 

Although in the case of quantum states the statistical properties differ from the commutative setting, needing whole new extension from the classical counterpart.
To be able to clearly compare its results with phenomena seen in classical systems, considered as having complex dynamical properties.
One could use statistics of local measurements on ensembles of quantum states $\rho$ in composite Hilbert spaces, $\mathcal{H}^{\otimes k},\,k>1$, to measure the same type of hierarchical information.
% the same procedure, if a set of POVMs are defined to bring the states, from their original Hilbert space to the simplex.
% And using the hierarchical information to measure correlations in this procedure.

In this setting it is shown that projective measurements ($\mathrm{PM}=\{P_i:\,P_i^2=P_i\}$) are more suitable due to their repeatability property and minimality assumptions, which exclude projective-operator valued measurements.
% For this purpose, projective measurements ($\mathrm{PM}=\{P_i:\,P_i^2=P_i\}$) are more suitable in the sense due to their repeatability property and due to minimality assumptions.
% Since the superset of positive-operator valued measurements (POVM) can be realized with an ancilla system together with unitary evolution on the composite system ancilla$\otimes$state, followed by a projective measurement on the ancilla system.
% Besides, PMs are maximal in the sense that no further information can be obtained by further measurements with the same observable over the state.
% Following the Gleason's theorem and thus using a von Neumann-Lüders set of projectors, that means projection onto a complete orthogonal set of subspaces; leads to a minimization on the statistical bias incurred from a bad choice in the set $\p$.
% Apart from natural restrictions as completeness $\sum_iP_i=\mathbb{I}$.
Following the Gleason's theorem and thus using a von Neumann-Lüders set of local projectors, this leads to a minimization on the statistical biasing effects incurred from a bad choice in the set $\mathrm{PM}$.
Also fulfilling natural restrictions, as completeness $\sum_iP_i=\mathbb{I}$,
the set of projectors formed using the computational basis $\p=\{P_i:\,P_i=\outm{i},\,i=0,1\}$ is one of the optimal ones within those restrictions.
% And also turns out to be state-independent in this setting, thus useful for the comparison possibility between the classical hierarchical information of distinct quantum states, via the use of the same set of measurements.

% Being left the only restrictions on the set of PMs are 
% The choice should then be state dependent, 
% This maximization can be done by means of optimizing the Fidelity between each projector and the state, feature possible due to quantum theory's self-duality.

% For that, the subset of von Neumann measurements, projective, is a priori chosen due to its repeatability property.
% Due to the fact that they surely leave the quantum state on a definite state after the measurement.
% 
% In this reduced set, of projective measurements, it is done an optimization to require maximal dintinguishability. 
% 
% $$\{P_i\}=min$$

%%%%%%%%%%%%%%%%%%%%%%%%%%%%%%%%%%%%%%%%%%%%%%%%%%%%%%%%%%%%%%%%%%%%%%%%%%%%%%%%%%%%%%%%%%%%%%%%%%%%%%%%%%%%%%%%%%%%%%%%%%%%%%%%%%%%%%%%%%%%%%%%%%%%%%%%%%%%%%%%%%%%%%%%%%%%

In order to assess the first higher hierarchical order $\I{3}$, two tripartite mixed state families are chosen.
For tripartite states it is known that there are only two classes of genuinely tri-party maximally entangled states, these are
$\ket{\mathrm{GHZ}}=\frac{\ket{000}+\ket{111}}{\sqrt{2}},$
the GHZ (Greenberger-Horne-Zeilinger) state and
$\ket{\mathrm{W}}=\frac{\ket{001}+\ket{010}+\ket{100}}{\sqrt{3}},$
the W-state.
Mixed-state families can then be build as their convex combination with the maximally mixed state $\mathbb{I}_2^{\otimes3}\in\mathcal{H}_2\otimes\mathcal{H}_2\otimes\mathcal{H}_2$:
$\varrho_{\mathrm{GHZ}}=\alpha\outm{GHZ}+\frac{1-\alpha}{8}\mathbb{I}_2^{\otimes3},\;\;
\varrho_{\mathrm{W}}=\alpha\outm{W}+\frac{1-\alpha}{8}\mathbb{I}_2^{\otimes3},$
respectively, and with $0\leq\alpha\leq1$.
% The use of a tripartite state is important on this quantification.
Naturally, every measured qubit takes the role of a system's fundamental unit, and then hierarchical correlations of order up to $k=3$ can be assessed, among those units.
% One can thus observe the behavior of $\I{3}$, which is the first complexity index in the interaction structures measurement.
% Because it describes the first high order interactions $k>2$ \cite{Kahle2009}.

% \vspace{-.6cm}

% \begin{figure}
  \begin{minipage}[t]{0.5\textwidth}
    \includegraphics[width=\textwidth,height=12cm]{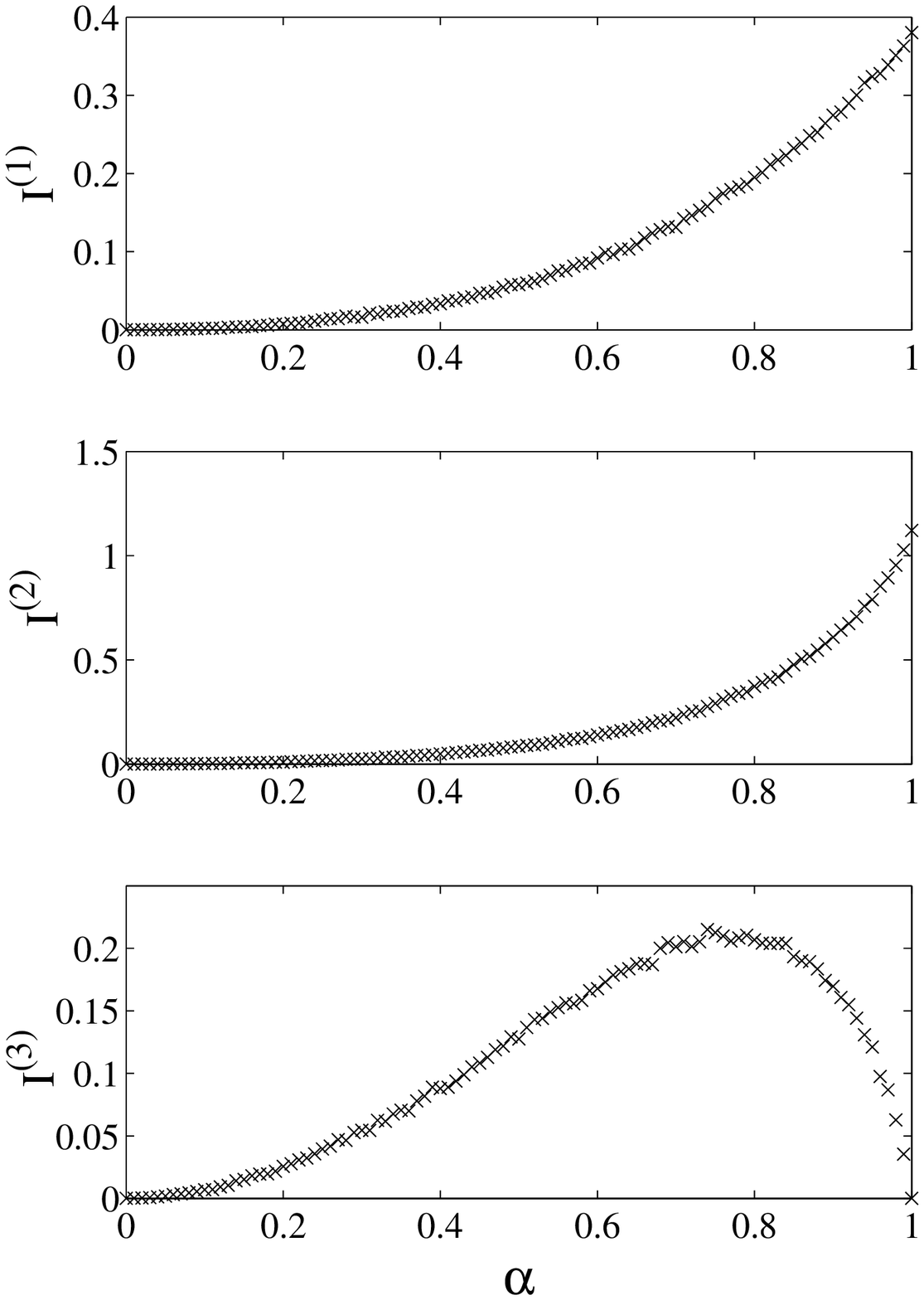}
\vspace{-1.3cm}
\captionof{figure}{Hierarchical information $\I{k}$ of $\varrho_{\mathrm{W}}$}
  \label{fig:iw}
  \end{minipage}
  \begin{minipage}[t]{0.5\textwidth}
    \includegraphics[width=\textwidth,height=12cm]{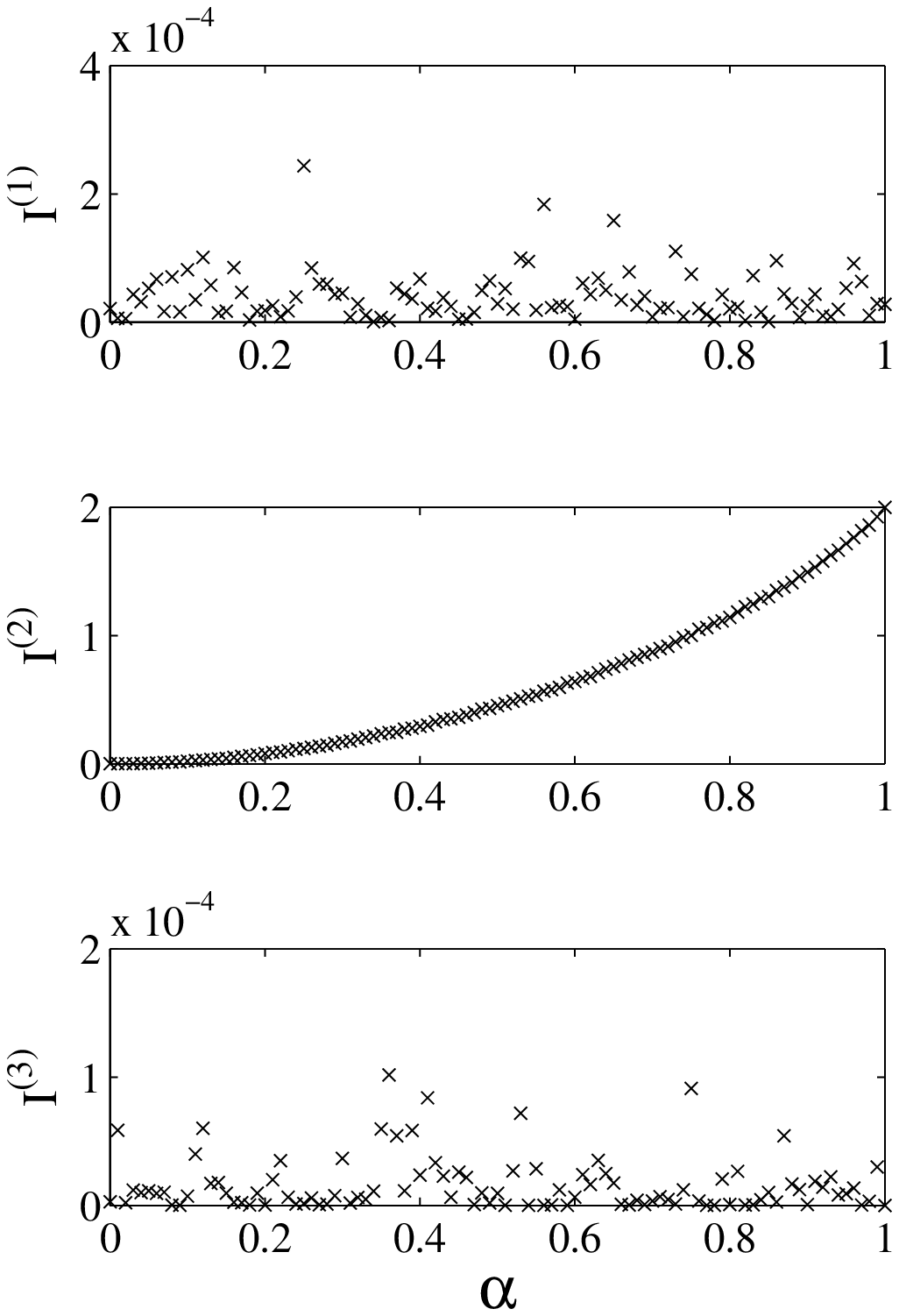}
\vspace{-1.3cm}
    \captionof{figure}{Hierarchical information $\I{k}$ of $\varrho_{\mathrm{GHZ}}$}
    \label{fig:ighz}
  \end{minipage}
% \vspace{.1cm}

Figures \ref{fig:iw} and \ref{fig:ighz} show the hierarchical information of the families of mixed states $\varrho_{\mathrm{W}}$ and $\varrho_{\mathrm{GHZ}}$, respectively, according to the chosen set of projective measurements $\mathrm{PM}$.
When $\alpha =0$, both mixed-state families are the maximally mixed state in $\mathcal{H}_2\otimes\mathcal{H}_2\otimes\mathcal{H}_2$.
But as the $\alpha -$values increase, one can notably see how the hierarchical information starts to build up differently.
Showing the effect of the distinct kind of entanglement between those families, since this is the only difference between the set-ups.

While on one hand the statistical behavior related to the family $\varrho_{\mathrm{GHZ}}$ can be successfully reduced to a pair interaction model, which is fully described by $\I{2}$ \cite{Kahle2009}.
On the other, the family $\varrho_{\mathrm{W}}$ shows up a more intricate type of hierarchical correlations.
In which the triple-wise correlations $\I{3}$ have a non-monotonically behavior with a maximum before the pure state entanglement takes part.
Albeit its lower order ones $\I{1,2}$ present monotonicity, and especially $\I{1}$, which shows that a measurement only on one qubit on the W states cannot uniquely determine the state of the other qubits, whereas this happens on the GHZ type of entanglement. Being also interesting to point out that the behavior of $\I{k}$ in $\varrho_{\mathrm{GHZ}}$ is closely related to statistics of classical synchronized dynamics, while $\I{k}$ in $\varrho_{\mathrm{W}}$ to statistics of classical complex systems dynamics.

\vspace{-.5cm}

\end{document}